\pdfoutput=1
\RequirePackage{ifpdf}
\ifpdf 
\documentclass[pdftex]{sigma}
\else
\documentclass{sigma}
\fi

\DeclareMathOperator\Artanh{artanh}

\begin{document}

\newcommand{\arXivNumber}{2103.01732}

\renewcommand{\PaperNumber}{057}

\FirstPageHeading

\ShortArticleName{Asymptotic Estimation for Eigenvalues in the Exponential Potential}

\ArticleName{Asymptotic Estimation for Eigenvalues\\ in the Exponential Potential and for Zeros of $\boldsymbol{K_{{\rm i}\nu}(z)}$\\ with Respect to Order}

\Author{Yuri KRYNYTSKYI and Andrij ROVENCHAK}

\AuthorNameForHeading{Yu.~Krynytskyi and A.~Rovenchak}

\Address{Department for Theoretical Physics, Ivan Franko National University of Lviv, Ukraine}

\Email{\href{mailto:yurikryn@gmail.com}{yurikryn@gmail.com}, \href{mailto:andrij.rovenchak@gmail.com}{andrij.rovenchak@gmail.com}}

\URLaddress{\url{https://physics.lnu.edu.ua/en/employee/krynytskyi-yu},\newline
\hspace*{10.5mm}\url{https://physics.lnu.edu.ua/en/employee/rovenchak-a}}

\ArticleDates{Received May 15, 2021, in final form June 01, 2021; Published online June 10, 2021}

\Abstract{The paper presents the derivation of the asymptotic behavior of $\nu$-zeros of the modified Bessel function of imaginary order $K_{{\rm i}\nu}(z)$. This derivation is based on the quasiclassical treatment of the exponential potential on the positive half axis. The asymptotic expression for the $\nu$-zeros (zeros with respect to order) contains the Lambert~$W$ function, which is readily available in most computer algebra systems and numerical software packages. The use of this function provides much higher accuracy of the estimation comparing to known relations containing the logarithm, which is just the leading term of~$W(x)$ at large~$x$. Our result ensures accuracies sufficient for practical applications.}


\Classification{33C10; 81Q05; 81Q20}

\section{Introduction}
The present work originates from an attempt to analyze the accuracy of numerical computations of the energy eigenvalues in steep potentials, one of which is the exponential potential. Additionally to purely mathematical interest linking this problem to finding the modified Bessel function zeros, such potentials appear in several physical problems, including quantum wells in~semiconductors~\cite{Guo_etal:2015,Sun&Xiao:2020,Yesilgul_etal:2019} and various cosmological models~\cite{Guo&Zhang:2005,Kamali_etal:2020}.

Even though the mathematical formulation of the problem is quite straightforward and can be relatively simply reduced to well-known modified Bessel functions, advancing to practical applications appears unexpectedly problematic. To be specific, the eigenvalues in the exponential potential are expressed via zeros of the modified Bessel function of the second kind (known also as the Macdonald function) $K_{{\rm i}\nu}(z)$ of imaginary order. The computation of the $K_{{\rm i}\nu}(z)$ zeros with respect to order, known as $\nu$-zeros, is not readily implemented in modern software even though algorithms for this were proposed decades ago~\cite{Campbell:1984,Cochran&Hoffspiegel:1970}. Moreover, available asymptotic expansions for large zeros reported in the literature~\cite{Bagirova&Khanmamedov:2020,Cochran:1965,Magnus&Kotin:1960} provide rather inaccurate estimations not applicable for direct calculations. Our aim is to fill in this gap.

The paper is organized as follows. In Section~\ref{sec:Quasi}, the quasiclassical approximation to the quantization in the exponential potential is considered. The equivalence of this problem with the problem of finding zeros of the Bessel $K_{{\rm i}\nu}(z)$ function is used in Section~\ref{sec:Kzeros} to obtain the asymptotic estimation for these zeros. Numerical comparison of the obtained asymptotics and previously suggested expression for zeros is made in Section~\ref{sec:Num}. Brief discussion in Section~\ref{sec:Concl} concludes the paper.

\section{Quasiclassical approximation}\label{sec:Quasi}

Consider a potential given by
\begin{gather*}
U(x) = \begin{cases}
U_0{\rm e}^{2x/a} & \text{for}\ x>0,
\\
+\infty& \text{for}\ x\leq0.
\end{cases}
\end{gather*}
The Hamiltonian for $x>0$ thus reads:
\begin{gather*}
H = \frac{p^2}{2m} + U_0{\rm e}^{2x/a}.
\end{gather*}

The Bohr--Sommerfeld quantization condition is given by
\begin{gather}\label{eq:QCond}
\int\limits_{x_1}^{x_2}\sqrt{2m[E-U(x)]}\,{\rm d}x = \pi\hbar\bigg(n+\frac34\bigg),
\end{gather}
where $x_1$ and $x_2$ are the classical turning points and the 3/4 correction originates from 1/2 as a contribution due to the hard wall at $x_1=0$ and another 1/4 contribution from the turning point $x_2$~\cite{Bhaduri_etal:2006,Curtis&Ellis:2004,Migdal&Krainov:1969,Vakarchuk:2012}. The quantization condition thus reads
\begin{gather}
\sqrt{2m}\int\limits_0^{\frac{a}{2}\ln\frac{E}{U_0}} \sqrt{E-U_0{\rm e}^{2x/a}}\,{\rm d}x =\sqrt{2m}\,a\bigg(\sqrt{E}\Artanh\sqrt{1-\frac{U_0}{E}}-\sqrt{E-{U_0}}\bigg)\nonumber
\\ \hphantom{\sqrt{2m}\int\limits_0^{\frac{a}{2}\ln\frac{E}{U_0}} \sqrt{E-U_0{\rm e}^{2x/a}}\,{\rm d}x}
{}=\pi\hbar\bigg(n+\frac34\bigg),\label{eq:WKB1}
\end{gather}
where $\Artanh z$ denotes the principal value of the inverse hyperbolic tangent. To approach an analytical solution of this equation we will expand the functions of $E$ into series taking into account that values of $E$ are large. Simple transformations yield
\begin{gather}\label{eq:ln2e}
\sqrt{\frac{E}{U_0}} \ln\frac2e + \sqrt{\frac{E}{U_0}}\ln\sqrt{\frac{E}{U_0}} +\mathcal{O}
\bigg(\sqrt{\frac{U_0}{E}}\bigg) =
\frac{\pi\hbar\big(n+\frac34\big)}{\sqrt{2mU_0a^2}}.
\end{gather}
Solutions for the transcendental equation
\begin{gather*}
bX + X\ln X = P
\end{gather*}
are given by
\begin{gather*}
X=\frac{P}{W({\rm e}^b P)},
\end{gather*}
where $W(z)$ is the Lambert function being the solution to
\begin{gather*}
W{\rm e}^W = z.
\end{gather*}
{\samepage
Finally, for the spectrum we obtain
\begin{gather}\label{eq:En-asymp}
E_n = \frac{\hbar^2}{2m}\bigg[\frac{\pi}{a}\bigg(n+\frac34\bigg)\bigg]^2
\frac{1}{W^2\bigg(\displaystyle \frac2e\,\frac{\pi\hbar\big(n+\frac34\big)}{\sqrt{2mU_0a^2}}\bigg)}.
\end{gather}
Note that here, as in the quantization condition (\ref{eq:QCond}), $n$ enumerates the wavefunction nodes, i.e., starts from $n=0$ at the ground state.}

\begin{figure}[t!]
\centering
\includegraphics[scale=0.7]{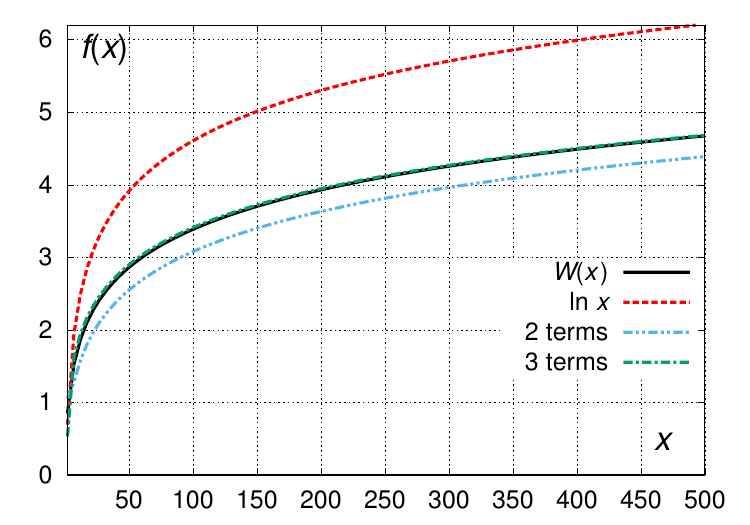}
\caption{Comparison of the Lambert $W$ function and its approximation by one, two, and three terms in the large $x$ series (\ref{eq:W-series}).}\label{fig:W-ln}
\end{figure}

For large arguments, the Lambert $W$ function can be expanded as a series in the following form~\cite{Corless_etal:1996}
\begin{align}\label{eq:W-series}
W(z) = \ln z - \ln\ln z + \frac{\ln\ln z}{\ln z} + \cdots.
\end{align}
The leading order of the spectrum for large $n$ is thus
\begin{align}\label{eq:En-asymp1}
E_n \propto \frac{n^2}{\ln^2 n}.
\end{align}
Series (\ref{eq:W-series}), however, converges very slowly and even the first two terms do not provide acceptable approximation to be used in practical calculations. For illustration, see Figure~\ref{fig:W-ln}. While the approximation with three terms seems satisfactory from the figure, this is an artifact of the vertical scale being too coarse. The accuracy of such an approximation is still far from good, as we will see in Section~\ref{sec:Num}.

{\sloppy
Presently, the Lambert $W$ is well implemented in computer algebra systems, such as Maple, Mathematica, Maxima, etc., and numerical software packages, including Perl, Python, R, \mbox{GNUPlot}, etc. It can be calculated using a fast converging iterative procedure with Newton's method~\cite{Johansson:2020}.

}

\section[Zeros of K i nu(z)]
{Zeros of $\boldsymbol{K_{{\rm i}\nu}(z)}$}\label{sec:Kzeros}

Expressions for eigenvalues of the Schr\"odinger equation with exponential potentials of the ${\rm e}^{-x}$ type have been long known to involve Bessel function zeros~\cite{Amore&Fernandez:2008,Bethe&Bacher:1936,Ma:1946,Sasaki&Znojil:2016}. A generalization of~the respective approaches for the positive sign in the exponential is straightforward and can be considered a textbook problem, cf.~\cite{Ahmed_etal:2018,Pisanty:2016}. Below we will briefly recall the derivation chain for consistency.

The stationary Schr\"odinger equation for our problem reads
\begin{gather*}
\bigg({-}\frac{\hbar^2}{2m}\frac{{\rm d}^2}{{\rm d}x^2} + U_0{\rm e}^{2x/a}\bigg)
\psi_n(x) = E_n\psi_n(x)
\end{gather*}
and the boundary conditions are
\begin{gather*}
\psi_n(0)=\psi_n(\infty) = 0.
\end{gather*}
Introducing the dimensionless variable $\xi=2x/a$ and using further $\hbar^2/\big(2ma^2\big)$ as the unit of energy we obtain the following equation
\begin{gather*}
\bigg({-}4\frac{{\rm d}^2}{{\rm d}\xi^2} + u{\rm e}^{\xi}\bigg)
\psi_n(\xi) = \varepsilon_n\psi_n(\xi),
\end{gather*}
where
\begin{gather*}
u=U_0 \frac{2ma^2}{\hbar^2},\qquad
\varepsilon_n = E_n\frac{2ma^2}{\hbar^2}.
\end{gather*}

Changing the variable $y=\sqrt{u}\,{\rm e}^{\xi/2}$ we ultimately arrive at
\begin{gather*}
y^2\, \frac{{\rm d}^2\psi_n(y)}{{\rm d}y^2} + y\, \frac{{\rm d}\psi_n(y)}{{\rm d}y} - (y^2 - \varepsilon_n)\psi_n(y) = 0.
\end{gather*}
The solutions of the above equation are given by the modified Bessel functions with the imaginary index ${\rm i}\nu = {\rm i}\sqrt{\varepsilon_n}$, namely
\begin{gather*}
\psi_n(y) = A_n I_{{\rm i}\nu}(y) + B_n K_{{\rm i}\nu}(y).
\end{gather*}
To satisfy the boundary condition $\psi_n(x=\infty) = \psi_n(y=\infty)=0$ we have to get rid of the function $I_{{\rm i}\nu}(y)$ that tends to infinity as $y\to\infty$. The wave functions are thus
\begin{gather*}
\psi_n(y) = B_n K_{{\rm i}\nu}(y).
\end{gather*}
Another boundary condition, $\psi_n(x=0) = \psi_n(y=\sqrt{u})=0$ yields
\begin{gather*}
K_{{\rm i}\nu}(\sqrt{u}) = 0.
\end{gather*}
The eigenvalues $\varepsilon_n$ are thus defined by the imaginary zeros ${\rm i}\nu_n$ of $K_{{\rm i}\nu}(\sqrt{u})$:
\begin{align*}
K_{{\rm i}\sqrt{\varepsilon_n}}(\sqrt{u}) = 0.
\end{align*}

\begin{figure}[h!]
\centering
\includegraphics[scale=0.7]{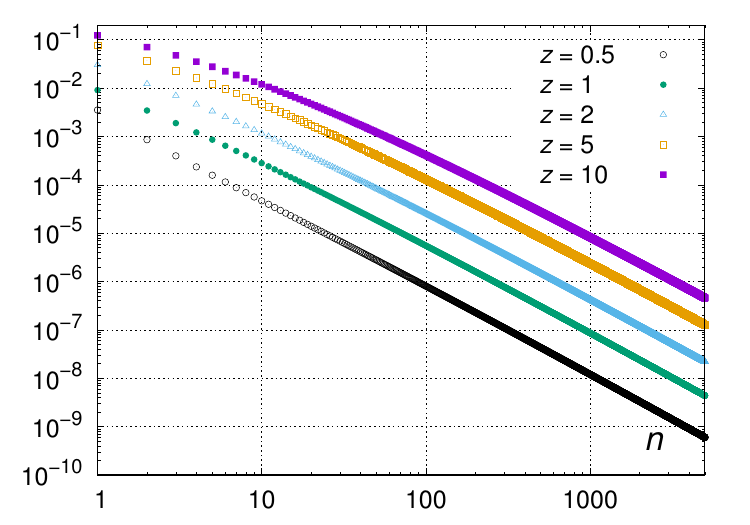}
\caption{Relative error $|\nu_n^{\rm asymp}/\nu_n-1|$ of the asymptotic expression (\ref{eq:Kzeros-asymp}) for zeros $\nu_n$ of $K_{{\rm i}\nu}(z)$.}\label{fig:Kerror}
\end{figure}

Simple expressions for $\nu_n$ are not known. Implicit expressions for them involve, in particular, Airy function zeros~\cite{Balogh:1967,Dunster:1990,Ferreira&Sesma:2008} while the asymptotic behavior of $\nu_n$ as $n\to\infty$ is known only in rather rough estimations~\cite{Bagirova&Khanmamedov:2020,Cochran:1965,Magnus&Kotin:1960} being weakly convergent due to the $\ln n$ and $\ln\ln n$ dependences. From our asymptotic estimation (\ref{eq:En-asymp}) for eigenvalues, shifting $n\to n-1$ to start counting zeros from unity, we have the following asymptotic behavior of the zeros
of $K_{{\rm i}\nu}(z)$:
\begin{gather}\label{eq:Kzeros-asymp}
\nu_n^{\rm asymp} =
\frac{\pi\big(n-\frac14\big)}{W\bigg(\displaystyle\frac{2\pi\big(n-\frac14\big)}{{\rm e}\,z}\bigg)}
\bigg[1+\mathcal{O}\bigg(\frac{\ln n}{n^2}\bigg)\bigg],
\qquad n= 1,2,3,\ldots\, .
\end{gather}
As we will see in the next section, this expression gives values quite close to the real ones even for small $n$. For instance, already the first zero of $K_{{\rm i}\nu}(1)$ is approximated with the relative error less than 1\%. For $K_{{\rm i}\nu}(2)$, the first zero has the relative error of 3\% dropping below 1\% already at the third zero. The accuracy of the approach is demonstrated in Figure~\ref{fig:Kerror}. The remainder term can be estimated from (\ref{eq:ln2e}).

\section[Comparison of asymptotic expressions for K i nu(z) zeros]
{Comparison of asymptotic expressions for $\boldsymbol{K_{{\rm i}\nu}(z)}$ zeros}\label{sec:Num}

\begin{figure}[h!]
\centering
\includegraphics[scale=0.7]{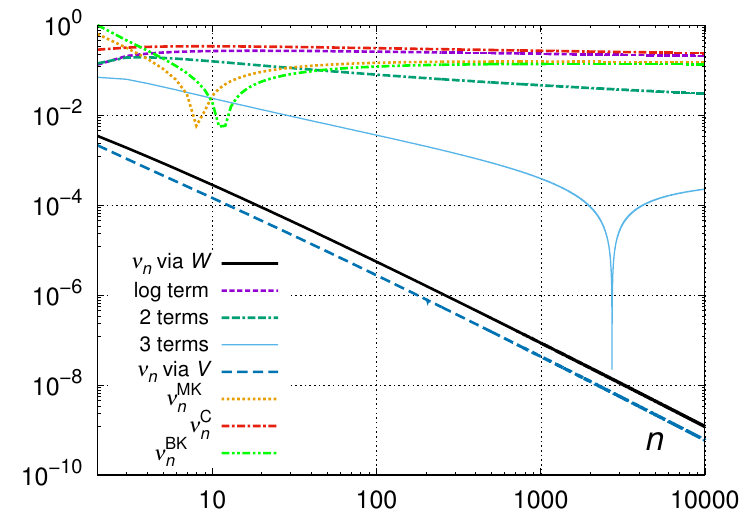}
\caption{Relative errors $|\nu_n^{\rm asymp}/\nu_n-1|$ of several asymptotic expressions (\ref{eq:Kzeros-asymp}) (including expansion~(\ref{eq:W-series})), (\ref{eq:Kzeros-asymp-compare}), and (\ref{eq:V-def})--(\ref{eq:nun-V}) for zeros $\nu_n$ of $K_{{\rm i}\nu}(1)$.}\label{fig:Kerror-compare}
\end{figure}

Zeros of $K_{\nu}$ coincide with those of the Hankel function $H^{(1)}_{\nu}$ in view of the relation~\cite{Abramowitz&Stegun:1972,Ferreira&Sesma:2008}
\begin{align*}
K_{\nu}(z) = \frac{{\rm i}\pi}{2}{\rm e}^{{\rm i}\pi\nu/2} H^{(1)}_{\nu}\big(z{\rm e}^{{\rm i}\pi/2}\big).
\end{align*}
So, one can use equally results for $K_\nu$ and $H^{(1)}_{\nu}$ available in the literature. Some closed-form asymptotic results can be obtained from both classical~\cite{Cochran:1965,Magnus&Kotin:1960} and new~\cite{Bagirova&Khanmamedov:2020} works. They are as follows
\begin{subequations}\label{eq:Kzeros-asymp-compare}
\begin{alignat}{3}
&\textrm{paper~\cite{Magnus&Kotin:1960}:}
\qquad &&\nu_n^{\rm MK} =
\frac{\pi\big(n+\frac14)}{\ln\bigg(\displaystyle\frac{\pi\big(n+\frac14\big)}{{\rm e}\,z}\bigg)}
\bigg[1+\mathcal{O}\bigg(\frac{\ln\ln n}{\ln n}\bigg)\bigg];&
\\
&\textrm{paper~\cite{Cochran:1965}:}
\qquad &&\nu_n^{\rm C} =
\frac{\pi n}{\ln\left(\displaystyle\frac{3\pi n}{{\rm e}\,z}\right)}
\bigg[1+\mathcal{O}\bigg(\frac{\ln\ln n}{\ln n}\bigg)\bigg];&
\\
&\textrm{paper~\cite{Bagirova&Khanmamedov:2020}:}
\qquad && \nu_n^{\rm BK} \sim
\frac{\pi n}{\ln n}.&
\end{alignat}
\end{subequations}
In Figure~\ref{fig:Kerror-compare} we compare the above asymptotics with our result expressed via the Lambert $W$ function (\ref{eq:Kzeros-asymp}) and its expansion (\ref{eq:W-series}) with one, two, and three terms.

As we can see, keeping the logarithmic term only cannot provide sufficient accuracy even for as large $\nu$-zero orders as $n\simeq 10^3$--$10^4$. The situation slightly improves with three terms in~expansion (\ref{eq:W-series}). The asymptotic containing exact $W(x)$ values yields the best result.

Curiously, the function $V(x)$ solving
\begin{gather}\label{eq:V-def}
\bigg(\Artanh\sqrt{1-\frac{1}{V}} - \sqrt{1-\frac{1}{V}}\bigg)V = x,
\end{gather}
that is, the exact quantization condition (\ref{eq:WKB1}), gives for the $\nu$-zeros of $K_{{\rm i}\nu}(z)$
\begin{align}\label{eq:nun-V}
\nu_n = zV\bigg(\frac{\pi\big(n-\frac14\big)}{z}\bigg)
\end{align}
that provides only a slightly better extimation comparing to equation~(\ref{eq:Kzeros-asymp}), as can be seen from Figure~\ref{fig:Kerror-compare}. The use of a similar function was suggested in~\cite{Ahmed_etal:2018}.

\section{Discussion}\label{sec:Concl}

One can show that for the $x^k$ potential on the positive half-axis the leading order of the eigenvalues is $E_n\propto n^{2k/(2+k)}$~\cite{Vakarchuk:2012}. Obviously, in the limit of $k\to\infty$ the energy levels are proportional to~$n^2$ corresponding to a particle in a box (the infinite square well). Considering the first correction to this dependence and the asymptotic relation (\ref{eq:En-asymp1}) we have
\begin{gather*}
E_n\propto \begin{cases}
n^2\,n^{-4/k} &\textrm{for}\ U(x)\propto x^k, \vspace{1mm}\\
\displaystyle\frac{n^2}{\ln^2 n} &\textrm{for}\ U(x)\propto {\rm e}^{2x/a}.
\end{cases}
\end{gather*}
This agrees with the fact that the exponential potential at large $x$ is steeper than \textit{any} power-law dependence $x^k$ hence the deviation from the square box $n^2$ for the exponential potential should be weaker that \textit{any} power $n^{\alpha\to -0}$ implying a logarithmic law.

The obtained asymptotic expressions for energy levels and directly related $\nu$-zeros of the modified Bessel functions contain the Lambert $W$ function. Nowadays, the calculation of this function is quite an easy task for most software so the appearance of $W(x)$ should not repel one from using these asymptotics. As we have shown, the substitution of the Lambert function with its leading logarithmic term cannot provide satisfactory numerical accuracy even for rather high energy levels $E_n$ or zeros $\nu_n$ for $n\sim 10^3$--$10^4$.

We should remark that preserving the 3/4 correction in the quantization condition (yielding the $-1/4$ shift in the expression for $\nu$-zeros) is essential for the accuracy of our approach not only for low values of $n$ but in the whole domain $n>0$. This correction allows keeping the relative error of the asymptotics as low as $\mathcal{O}\big(\frac{\ln n}{n^2}\big)$.

\looseness=1
To summarize, we propose an asymptotic estimation for the eigenvalues in the exponential potential and for the $\nu$-zeros of the modified Bessel function of the second kind with and imaginary order, which are of sufficient accuracy to be used in practical calculations. This estimations are~expressed in particular via the Lambert $W$ function being presently well implemented in~modern software. Further attempts to improve the approach based on the numerical solution of~the equation for the quasiclassical quantization condition only lead to an insignificant increase in~the accuracy. Our asymptotic is also useful as an initial guess for the numerical computation of~$\nu$-zeros of a given order as well as and estimation for distances between consecutive zeros.

{\sloppy
For convenience, we uploaded to the online repository at \url{https://doi.org/10.5281/zenodo.4573305} the data for first 10\,000 zeros for several values of $z$ together with their asymptotic esti\-mations according to our formula.

}\vspace{-2mm}

\pdfbookmark[1]{References}{ref}
\LastPageEnding


\begin{thebibliography}{10}
\footnotesize\itemsep=-1.7pt

\bibitem{Abramowitz&Stegun:1972}
Abramowitz M., Stegun I.A. (Editors), Handbook of mathematical functions, with
 formulas, graphs, and mathematical tables, Dover Publications, Inc., New
 York, 1966.

\bibitem{Ahmed_etal:2018}
Ahmed Z., Ghosh D., Kumar S., Turumella N., Solvable models of an open well and
 a bottomless barrier: one-dimensional exponential potentials, \href{https://doi.org/10.1088/1361-6404/aa8c0c}{\textit{Eur.~J.
 Phys.}} \textbf{39} (2018), 025404, 10~pages, \href{https://arxiv.org/abs/1706.05275}{arXiv:1706.05275}.

\bibitem{Amore&Fernandez:2008}
Amore P., Fern\'andez M.F., Accurate calculation of the complex eigenvalues of the
 {S}chr\"odinger equation with an exponential potential, \href{https://doi.org/10.1016/j.physleta.2008.01.053}{\textit{Phys.
 Lett.~A}} \textbf{372} (2008), 3149--3152, \href{https://arxiv.org/abs/0712.3375}{arXiv:0712.3375}.

\bibitem{Bagirova&Khanmamedov:2020}
Bagirova S.M., Khanmamedov A.K., On zeros of the modified {B}essel function of
 the second kind, \href{https://doi.org/10.1134/S0965542520050048}{\textit{Comput. Math. Math. Phys.}} \textbf{60} (2020),
 817--820.

\bibitem{Balogh:1967}
Balogh C.B., Asymptotic expansions of the modified {B}essel function of the
 third kind of imaginary order, \href{https://doi.org/10.1137/0115114}{\textit{SIAM~J. Appl. Math.}} \textbf{15}
 (1967), 1315--1323.

\bibitem{Bethe&Bacher:1936}
Bethe H.A., Bacher R.F., Nuclear physics~{A}. {S}tationary states of nuclei,
 \href{https://doi.org/10.1103/RevModPhys.8.82}{\textit{Rev. Mod. Phys.}} \textbf{8} (1936), 82--229.

\bibitem{Bhaduri_etal:2006}
Bhaduri R.K., Sprung D.W.L., Suzuki A., When is the lowest order {WKB}
 quantization exact?, \href{https://doi.org/10.1139/p06-024}{\textit{Can.~J. Phys.}} \textbf{84} (2006), 573--581,
 \href{https://arxiv.org/abs/gr-qc/0508107}{arXiv:gr-qc/0508107}.

\bibitem{Campbell:1984}
Campbell J., Determination of $\nu$-zeros of {H}ankel functions,
 \href{https://doi.org/10.1016/0010-4655(84)90095-x}{\textit{Comput. Phys. Comm.}} \textbf{32} (1984), 333--339.

\bibitem{Cochran:1965}
Cochran J.A., The zeros of {H}ankel functions as functions of their order,
 \href{https://doi.org/10.1007/BF01436080}{\textit{Numer. Math.}} \textbf{7} (1965), 238--250.

\bibitem{Cochran&Hoffspiegel:1970}
Cochran J.A., Hoffspiegel J.N., Numerical techniques for finding {$\nu $}-zeros
 of {H}ankel functions, \href{https://doi.org/10.2307/2004488}{\textit{Math. Comp.}} \textbf{24} (1970), 413--422.

\bibitem{Corless_etal:1996}
Corless R.M., Gonnet G.H., Hare D.E.G., Jeffrey D.J., Knuth D.E., On the
 {L}ambert {$W$} function, \href{https://doi.org/10.1007/BF02124750}{\textit{Adv. Comput. Math.}} \textbf{5} (1996),
 329--359.

\bibitem{Curtis&Ellis:2004}
Curtis L.J., Ellis D.G., Use of the {E}instein--{B}rilloui--{K}eller action
 quantization, \href{https://doi.org/10.1119/1.1768554}{\textit{Amer.~J. Phys.}} \textbf{72} (2004), 1521--1523.

\bibitem{Dunster:1990}
Dunster T.M., Bessel functions of purely imaginary order, with an application
 to second-order linear differential equations having a large parameter,
 \href{https://doi.org/10.1137/0521055}{\textit{SIAM~J. Math. Anal.}} \textbf{21} (1990), 995--1018.

\bibitem{Ferreira&Sesma:2008}
Ferreira E.M., Sesma J., Zeros of the {M}acdonald function of complex order,
 \href{https://doi.org/10.1016/j.cam.2006.11.014}{\textit{J.~Comput. Appl. Math.}} \textbf{211} (2008), 223--231,
 \href{https://arxiv.org/abs/math.CA/0607471}{arXiv:math.CA/0607471}.

\bibitem{Guo_etal:2015}
Guo K.-X., Xiao B., Zhou Y., Zhang Z., Polaron effects on the third-harmonic
 generation in asymmetrical semi-exponential quantum wells, \href{https://doi.org/10.1088/2040-8978/17/3/035505}{\textit{J.~Optics}}
 \textbf{17} (2015), 035505, 6~pages.

\bibitem{Guo&Zhang:2005}
Guo Z.-K., Zhang Y.-Z., Interacting phantom energy, \href{https://doi.org/10.1103/physrevd.71.023501}{\textit{Phys. Rev.~D}}
 \textbf{71} (2005), 023501, 5~pages, \href{https://arxiv.org/abs/1910.06796}{arXiv:1910.06796}.

\bibitem{Johansson:2020}
Johansson F., Computing the {L}ambert {$W$} function in arbitrary-precision
 complex interval arithmetic, \href{https://doi.org/10.1007/s11075-019-00678-x}{\textit{Numer. Algorithms}} \textbf{83} (2020),
 221--242, \href{https://arxiv.org/abs/1705.03266}{arXiv:1705.03266}.

\bibitem{Kamali_etal:2020}
Kamali V., Motaharfar M., Ramos R.O., Warm brane inflation with an exponential
 potential: a consistent realization away from the swampland, \href{https://doi.org/10.1103/physrevd.101.023535}{\textit{Phys.
 Rev.~D}} \textbf{101} (2020), 023535, 13~pages, \href{https://arxiv.org/abs/1910.06796}{arXiv:1910.06796}.

\bibitem{Ma:1946}
Ma S.T., Redundant zeros in the discrete energy spectra in {H}eisenberg's
 theory of characteristic matrix, \href{https://link.aps.org/doi/10.1103/PhysRev.69.668}{\textit{Phys. Rev.}} \textbf{69} (1946),
 668--668.

\bibitem{Magnus&Kotin:1960}
Magnus W., Kotin L., The zeros of the {H}ankel function as a function of its
 order, \href{https://doi.org/10.1007/BF01386226}{\textit{Numer. Math.}} \textbf{2} (1960), 228--244.

\bibitem{Migdal&Krainov:1969}
Migdal A.B., Krainov V., Approximation methods in quantum mechanics,
 W.A.~Benjamin, Inc., New York~-- Amsterdam, 1969.

\bibitem{Pisanty:2016}
Pisanty E., {A}nswer to: {E}igenvalues and eigenfunctions of the exponential
 potential $V(x)=\exp(|x|)$, 2016, available at
 \url{https://physics.stackexchange.com/questions/47128/eigenvalues-and-eigenfunctions-of-the-exponential-potential-vx-expx}.

\bibitem{Sasaki&Znojil:2016}
Sasaki R., Znojil M., One-dimensional {S}chr\"odinger equation with
 non-analytic potential {$V(x)=-g^2\exp (-|x|)$} and its exact
 {B}essel-function solvability, \href{https://doi.org/10.1088/1751-8113/49/44/445303}{\textit{J.~Phys.~A: Math. Theor.}} \textbf{49}
 (2016), 445303, 12~pages, \href{https://arxiv.org/abs/1605.07310}{arXiv:1605.07310}.

\bibitem{Sun&Xiao:2020}
Sun Y., Xiao J.-L., Coherence effects of the strongly-coupled optical
 polaron-level qubit in a quantum well with asymmetrical semi-exponential
 potential, \href{https://doi.org/10.1016/j.spmi.2020.106617}{\textit{Superlattices Microstruct.}} \textbf{145} (2020), 106617,
 7~pages.

\bibitem{Vakarchuk:2012}
Vakarchuk I., Quantum mechanics, Lviv University Press, Lviv, 2012.

\bibitem{Yesilgul_etal:2019}
Yesilgul U., Ungan F., Sakiroglu S., Sari H., Kasapoglu E., S\"okmen I.,
 Nonlinear optical properties of a~semi-exponential quantum wells: effect of
 high-frequency intense laser field, \href{https://doi.org/10.1016/j.ijleo.2019.03.126}{\textit{Optik}} \textbf{185} (2019),
 311--316.

\end{thebibliography}
\end{document}